\documentclass[prc,twocolumn,nofootinbib,showpacs,showkeys,floatfix,preprintnumbers,byrevtex,superscriptaddress]{revtex4}

\usepackage[latin1]{inputenc}
\usepackage[T1]{fontenc}
\usepackage{epsfig}
\usepackage{bm}
\usepackage{amsmath,amsfonts,amssymb,amscd,amsxtra,amsthm}
\usepackage{graphicx}
\usepackage[usenames]{color} 
\usepackage{bbm}
\usepackage{multirow}
\newcommand{\tr}{\mathrm{tr}}

\newcommand{\bra}{\langle}
\newcommand{\ket}{\rangle}
\newcommand{\qucond}{\bra\bar{q}q\ket}
\newcommand{\glcond}{\left\bra\frac{\alpha_s}{\pi}G^2\right\ket}

\newcommand{\ud}{\mathrm{d}}
\newcommand{\timeop}{\mathcal{T}}

\begin{document}
\preprint{TUM-T39-10-05}
\title{Vector mesons at finite temperature and QCD sum rules}

\author{Youngshin Kwon}
\email{ykwon@ph.tum.de}
\affiliation{Physik-Department, Technische Universit\"at M\"unchen,\\D-85747 Garching, Germany}

\author{Chihiro Sasaki}
\email{sasaki@fias.uni-frankfurt.de}
\affiliation{Frankfurt Institute for Advanced Studies, D-60438 Frankfurt am Main, Germany}

\author{Wolfram Weise\footnote{Temporary address: Yukawa Institute of Theoretical Physics,\\University of Kyoto, Japan}}
\email{weise@ph.tum.de}
\affiliation{Physik-Department, Technische Universit\"at M\"unchen,\\D-85747 Garching, Germany}


\begin{abstract}
Finite energy sum rules for vector and axial-vector currents are derived in a thermal medium to provide constraints for the spectral behavior of $\rho$ and $a_1$ mesons at nonvanishing temperature and hence to study the tendency toward chiral symmetry restoration. The parity-mixing ansatz for the $\rho$ and $a_1$ spectra, including finite widths, is investigated as a function of temperature. Characteristic differences between vector and axial-vector channels are discussed with regard to the implementation of the chiral-symmetry-breaking scale, $4\pi$ times the pion decay constant, in the sum rule approach.  
\end{abstract}


\pacs{11.30.Qc, 11.30.Rd, 11.55.Hx, 14.40.Be}

\keywords{Vector meson, chiral symmetry, finite temperature and QCD sum rules}

\maketitle

\section{Introduction}
The chiral $SU(N_f)_L\times SU(N_f)_R$ symmetry of the QCD Lagrangian with $N_f$ (massless) quark flavors is known to be broken down spontaneously to $SU(N_f)_V$ in the
nonperturbative QCD vacuum. With inclusion of explicit chiral symmetry breaking by the nonvanishing quark masses, this symmetry-breaking pattern is visibly manifest in the spectrum of the low-mass pseudoscalar meson octet for $N_f = 3$. This octet figures as the massless Goldstone bosons of spontaneously broken chiral $SU(3)_L\times SU(3)_R$ symmetry in the limit of massless $u$, $d$, and $s$ quarks.  An equivalent but not directly observable manifestation of spontaneous chiral symmetry breaking (S$\chi$SB) is the appearance of a nonzero quark condensate. Its observable counterpart is the pion decay constant, $f_\pi$, which acts as an order parameter for S$\chi$SB and defines a chiral-symmetry-breaking scale, $4\pi f_\pi \sim 1$ GeV, briefly referred to as the chiral scale. Further evidence for S$\chi$SB is the absence of parity doublets in the lower mass area of the hadron spectrum, an important example being the large mass gap between the $\rho$ $(J^P = 1^-)$ and $a_1$ $(J^P = 1^+)$ mesons. Thus it is well established that at zero temperature, QCD is in the Nambu-Goldstone realization of chiral symmetry. Were it in the Wigner-Weyl realization with a trivial vacuum and parity doublets, the $\rho$ and $a_1$ would be degenerate.  One expects this latter situation to be reached as the temperature is raised toward a critical value, $T_c \sim 0.2$ GeV, for the chiral transition. Around this point and beyond, the $\rho$ and $a_1$ are expected to melt into the quark-antiquark continuum.

Lattice QCD results~\cite{Cheng:2006qk, Aoki:2006br} do indicate that the spontaneously broken chiral
symmetry is restored at high temperature. The spectral functions of chiral partners such as the (isovector) vector and and axial-vector mesons are then supposed to become degenerate.  In this context, concerning the in-medium behavior of vector mesons,  measurements of dileptons produced
in relativistic heavy-ion collisions have attracted great interest over the past decades (see, e.g., Ref. \cite{NA60}). Dileptons as well as photons are excellent probes carrying information about vector meson spectral functions at the high temperatures and densities reached in the collision zone. In particular, the $\rho$ meson, as the lowest dipole excitation of the QCD vacuum, has been in the focus
of such investigations.

The issue of in-medium hadronic properties persists as a basic theme ever since the suggestion of Brown and
Rho (BR)~\cite{Brown:1991kk} that hadron masses should decrease systematically when a hadronic medium undergoes the transition toward chiral symmetry restoration. The BR scaling hypothesis states that the dropping of hadron masses, with the exception of the pion mass, should go in parallel with the dropping of the pion decay constant $f_\pi$ (here the one related to the time component of the axial current) in hot and dense hadronic matter. In general, a sufficient condition for chiral symmetry restoration is that  the in-medium mass difference between the $\rho$ and $a_1$ mesons tends to zero, although it is not mandatory that the masses vanish individually. In practice, however, these mass-shift scenarios are overshadowed by strong spectral-broadening effects. The $\rho$ and $a_1$ resonances have large widths already in the vacuum. Their interactions with hadrons in the medium increase their decay widths significantly through collision broadening. Distinguishing ``mass shift'' from ``broadening'' scenarios is obviously not a meaningful issue once the spectral functions become very broad and do not show clear resonance structures. 

In a situation like this, it nevertheless still makes sense to perform a sum-rule analysis of such broad spectral distributions, focusing on their lowest moments and defining a mean mass through
the first moment of the in-medium spectral function. Such an analysis is made possible by identifying the characteristic chiral-symmetry-breaking scale, $\Lambda_{\mathrm{CSB}}\approx4\pi f_\pi$, with the continuum threshold separating the low-energy-resonance region from the high-energy continuum. Consider the example of the vector spectral function, to be specified in detail later:
\begin{equation}
    R_V(s)=R_V^{res}(s)\,\Theta(s_V-s)+R_V^{cont}\,\Theta(s-s_V)~,
\end{equation}
where $R_V^{res}(s)$ denotes the resonance part. This low-energy part is separated by the scale $s_V$ from the high-energy continuum part $R_V^{cont}$ determined by perturbative QCD. In vacuum, the identification $\sqrt{s_V}=4\pi f_\pi$ is supported by current algebra relations and spectral sum rules, as we point out later. The extension to the case of finite baryon density at zero temperature has been studied in our previous work~\cite{Kwon:2008vq} where it is demonstrated that the aforementioned scale analysis works with $\sqrt{s_V(\rho)} = 4\pi f_\pi^*(\rho)$ now interpreted in terms of the in-medium change of the chiral order parameter. 

Before turning to finite temperatures,  it is useful to recall Weinberg's sum rules ~\cite{Weinberg:1967kj}. These sum rules are entirely based on current algebra and establish rigorous relations for the difference between vector and axial-vector spectral functions:
\begin{eqnarray}
   \int^{\infty}_0\ud s\,\big[R_V(s)-R_A(s)\big]&=&0~,
    \label{eq:Wsr1}\\
   \int^{\infty}_0\ud s\,s\big[R_V(s)-R_A(s)\big]&=&0~.
    \label{eq:Wsr2}
\end{eqnarray}
Assume first the following schematic forms  for the vector and axial-vector spectral functions: 
\begin{equation}
 \begin{split}
    R_{V}(s)&=R_{V}^{res}(s)\,\Theta(s_{V}-s)+c_0\,\Theta(s-s_{V})~,\\
    R_{A}(s)&=R_{A}^{res}(s)\,\Theta(s_{A}-s)+c_0\,\Theta(s-s_{A})~,\\ 
\end{split}
\label{eq:specftn0}
\end{equation}
where $c_0$ multiplying the continuum parts of the spectral functions is determined by perturbative QCD $(c_0 = 3/2$ in leading order for the $\rho$ and $a_1$ channels). The equality of these vector and axial-vector continuum pieces is indicative of chiral symmetry restoration
at high energy beyond the scales $s_{V,A}$. For the resonant parts, choose for simplicity a zero-width ansatz (as realized
in the large-$N_c$ limit):
\begin{equation}
 \begin{split}
    R_V^{res}(s)&=12\pi^2\,f_V^2 m_V^2\,\delta(s-m_V^2)~,\\
    R_A^{res}(s)&=12\pi^2\,\big[f_\pi^2\,\delta(s-m_\pi^2)+f_A^2m_A^2\,\delta(s-m_A^2)\big]~,
 \end{split}
\label{eq:specftn0prime}
\end{equation}
where  $f_{V/A}$ are dimensionless vector and axial-vector couplings respectively. The $a_1$ spectral function includes a pion pole term with its residue determined by the pion decay constant $f_\pi$. In the chiral limit ($m_\pi\rightarrow 0$), and assuming equal continuum thresholds for the vector and axial-vector channels, $s_V = s_A$, the Weinberg sum rules (\ref{eq:Wsr1},\ref{eq:Wsr2}) read as follows:
\begin{equation}
 \begin{split}
    f_V^2m_V^2-f_A^2m_A^2&=f_\pi^2~,\\
    f_V^2m_V^4-f_A^2m_A^4&=0~.
 \end{split}
\label{eq:largeNWSR}
\end{equation}
These equations, together with the KSRF relation~\cite{Kawarabayashi:1966kd,Riazuddin:1966sw},
\begin{equation}
f_V^2m_V^2=2f_\pi^2~,
\label{eq:lKSFR}
\end{equation}
imply that the vacuum $a_1$ and $\rho$ masses satisfy the well-known relation 
\begin{equation}
m_A^2=2m_V^2~.
\label{eq:mass}
\end{equation} 

The Weinberg sum rules are based on the observation that chiral symmetry is restored in its Wigner-Weyl realization asymptotically, at high-energy scales, where QCD is perturbative. The duality between the resonant and asymptotic parts of the spectral functions derived from current correlators is one of the basic ideas of QCD sum rules, to which we now proceed. The present study constructs finite energy sum rules (FESR) for $\rho$ and $a_1$ meson at nonzero temperature and vanishing baryon chemical potential. We show how finite temperature FESRs can provide constraints for the pattern and trend toward chiral symmetry restoration. Realistic vacuum spectral functions are employed as input in comparison with schematic $\delta$ function spectra, and the $\rho$-$a_1$ parity-mixing scenario is used to describe temperature-dependent effects.

\section{Finite energy sum rules}

Following these preparations, the starting point is now the time-ordered current correlation function
\begin{equation}
    \Pi^{\mu\nu}(q)=i\int\ud^4x\,e^{iq\cdot x}\bra\timeop
    j^\mu(x)j^\nu(0)\ket_T,
        \label{CCcorelator}
\end{equation}
with the vector current $j_V^\mu(x)=\frac{1}{2}\left(\bar{u}\gamma^\mu u-\bar{d}\gamma^\mu d\right)$ and the axial-vector
current $j_A^\mu=\frac{1}{2}\left(\bar{u}\gamma^\mu\gamma_5u-\bar{d}\gamma^\mu\gamma_5d\right)$ carrying the quantum
numbers of $\rho$ and $a_1$ meson, respectively. The bracket $\bra\mathcal{O}\ket_T$ indicates the thermal expectation value of an operator $\mathcal{O}$,
\begin{equation}
    \bra\mathcal{O}\ket_T=\frac{\tr\,\mathcal{O}\,\exp\left(-H/T\right)}{\tr\,\exp\left(-H/T\right)},
        \label{eq:TherAV}
\end{equation}
where $H$ is the Hamiltonian.

In vacuum, the tensor correlation function (\ref{CCcorelator}) can be related to a single invariant correlator, $\Pi(q^2)=\frac{1}{3}g_{\mu\nu}\Pi^{\mu\nu}$. In a medium, the longitudinal and transverse parts of the correlator differ as a consequence of broken Lorentz invariance. In the rest frame of the medium and for the case in which the mesons have vanishing three-momentum $\mathbf{q}=0$, however, longitudinal and transverse correlation functions coincide and again are given as a single function $\Pi(\omega,\mathbf{q}=0)$. This correlator is written in the form of a twice-subtracted dispersion relation:
\begin{equation}
    \Pi(q^2)=\Pi(0)+\Pi^\prime(0)\,q^2+\frac{q^4}{\pi}\int{\ud s}\frac{\mathrm{Im}\Pi(s)}{s^2(s-q^2-i\epsilon)}.
\label{dispersion}
\end{equation}

On the other hand, the operator product expansion (OPE) is used to represent the correlator at large spacelike momentum, $q^2=-Q^2<0$:
\begin{equation}
    12\pi^2\Pi(q^2 = -Q^2)=-c_0Q^2\ln\left(\frac{Q^2}{\mu^2}\right)+c_1+\frac{c_2}{Q^2}+\frac{c_3}{Q^4}+\cdots~,
\label{ope}
\end{equation}
with the coefficients
\begin{equation}
  \begin{split}
    c_0&=\frac{3}{2}\left(1+\delta_N\right)~,\\
    c_1&=-\frac{9}{2}(m^2_u+m^2_d)~,\\
    c_2&=\frac{\pi^2}{2}\glcond_T\pm6\pi^2\left(m_u\bra\bar{u}u\ket_T+m_d\bra\bar{d}d\ket_T\right)~.
  \end{split}
\label{c_n}
\end{equation}
Here, $\delta^{}_N$ in $c_0$ denotes the radiative corrections in perturbative QCD. The explicit 
form of $\delta^{}_N$ up to order $\alpha_s^3(s)$ can be found in Ref.~\cite{Kwon:2008vq}. The difference between vector and axial-vector channels results to this order from the sign of the quark condensate term in $c_2$. In the chiral limit ($m_{u,d}\rightarrow 0$), $c_1$ and the second term of $c_2$ vanish. The first appearance of a difference between $\rho$ and $a_1$ spectral functions is then in the term proportional  to $c_3$ in the expansion (\ref{ope}). This term involves four-quark condensates, 
\begin{equation}
  \begin{split}
    c_3&=-6\pi^3\alpha_s\big[\bra(\bar{u}\gamma_\mu\gamma_5\lambda^au\mp\bar{d}\gamma_\mu\gamma_5\lambda^ad)^2\ket_T\\
    &\quad+\frac{2}{9}\bra(\bar{u}\gamma_\mu\lambda^au+\bar{d}\gamma_\mu\lambda^ad)\sum_{q=u,d,s}\bar{q}\gamma^\mu\lambda^aq\ket_T\big]~,
  \end{split}
\label{c_3}
\end{equation}
that are subject to large uncertainties. To evaluate these condensates, a factorization approximation is frequently used, assuming that intermediate states are saturated by the QCD ground state:
\begin{equation}
    \bra(\bar{q}\gamma_\mu\gamma_5\lambda^a{q})^2\ket=-\bra(\bar{q}\gamma_\mu\lambda^a{q})^2\ket
    =\frac{16}{9}\kappa\,\bra\bar{q}q\ket^2~,
\label{4q}
\end{equation}
with $\kappa$ introduced to parametrize the deviation from exact factorization ($\kappa=1$). 

However, the factorization approximation for the four-quark condensates is not sufficiently accurate for any quantitative considerations~\cite{Kwon:2008vq}. This basic uncertainty can be avoided by arranging sum rules in terms of moments of the spectral function and restricting the analysis to the lowest two moments:
\begin{eqnarray}
  \int^{s_0(T)}_0\ud s\,R(s,T)&=& \nonumber
  \\s_0(T)\,c_0 &+& c_1-12\pi^2\,\Pi(0)~,
  \label{0momsr}\\
  \int^{s_0(T)}_0\ud s\,s\,R(s,T)&=&\frac{s^2_0(T)}{2}\,c_0-c_2(T)~.
  \label{1momsr}
\end{eqnarray}
The dimensionless, temperature-dependent spectral function $R(s,T)$ stands generically for the vector or axial channel and is defined as
\begin{equation}
R(s,T)=-\frac{12\pi}{s}\mathrm{Im}\Pi(s,T)~. \nonumber
\end{equation}
Its high-energy continuum is separated from the low-energy part by the scale $s_0(T)$ that is expected to shift downward with increasing temperature. Again, this $s_0$ stands for the continuum
threshold scale in either vector or axial-vector channel. These scales, denoted in the following by $s_V(T)$ and $s_A(T)$, are generally different and to be determined by the detailed sum-rule analysis. The last term on the right-hand side of Eq.~(\ref{0momsr}) vanishes in the vector channel and represents the pion pole contribution in the axial-vector channel, with $\Pi(0)=f_\pi^2$ at $T=0$.

In the asymptotic region ($Q^2\to\infty$) where the OPE is valid, all nonperturbative scales appear in the form of power corrections to the perturbative calculations. These nonperturbative contributions are separated into the respective condensates. At low temperatures, it can be assumed that, apart from the scales $s_{V,A}(T)$, the $T$ dependence is only in the condensates.

\section{Evaluation of $\bra\mathcal{O}\ket_T$}
The evaluation of the $T$-dependent condensates follows the method employed previously in Ref.~\cite{Hatsuda:1992bv}. In Eq.~(\ref{eq:TherAV}), the vacuum state and the lowest excitations of the hadron gas, namely pions, are taken into account as eigenstates of the Hamiltonian to calculate the thermal expectation values at low temperatures:
\begin{equation}
    \bra\mathcal{O}\ket_T=\bra\mathcal{O}\ket_0+\sum^3_{a=1}\int\frac{\ud^3p}{2E(2\pi)^3}\,\bra\pi^a(p)|\mathcal{O}|\pi^a(p)\ket\, n^{}_B~,
        \label{eq:matrixT}
\end{equation}
where $n^{}_B=\left(e^{E/T}-1\right)^{-1}$ denotes Bose-Einstein distributions of thermal pions with $E^2=m_\pi^2+p^2$.
The pion matrix element in Eq.~(\ref{eq:matrixT}) can be evaluated in the soft pion limit:
\begin{equation}
    \bra\pi^a(p)|\mathcal{O}|\pi^a(p)\ket=-\frac{1}{f_\pi^2}\bra0|\left[\mathcal{Q}^a_5,\,\left[\mathcal{Q}^a_5,\,\mathcal{O}\right]\right]|0\ket+\cdots,
     \label{eq:softpion}
\end{equation}
where $\mathcal{Q}_5^a$ is the axial charge operator. Equations (\ref{eq:matrixT}) and (\ref{eq:softpion}), when applied to the scalar quark operator, $\bar{q}q$, give the leading order $T$ dependence of the chiral condensate,
\begin{equation}
    \qucond_T=\qucond_0\left[1-\frac{T^2}{8f_\pi^2}B_1\left(\frac{m_\pi}{T}\right)\right]~,
        \label{eq:LOqucondT}
\end{equation}
with
\begin{equation}
    B_1(x)=\frac{6}{\pi^2}\int^\infty_x \ud
    y\,\frac{\sqrt{y^2-x^2}}{e^y-1}~.
\end{equation}
Equation (\ref{eq:LOqucondT}) is the well-known result derived from chiral effective field theory~\cite{Gasser:1986vb,Gerber:1988tt,Kaiser:1999mt}.

The application of Eq.~(\ref{eq:matrixT}) to the gluon condensate makes use of the QCD trace anomaly,
\begin{equation}
    \Theta^\mu_\mu=-\frac{1}{8}\left(11-\frac{2}{3}N_f\right)\frac{\alpha_s}{\pi}G^2+\sum_q m_q\bar{q}q~,
\end{equation} 
to calculate the pionic matrix element of  the relevant gluon operator
\cite{Hatsuda:1990uw}. For $N_f=3$, the resulting temperature dependent gluon condensate becomes
\begin{equation}
    \glcond_T=\glcond_0-\frac{1}{9}m_\pi^2T^2B_1\left(\frac{m_\pi}{T}\right)~,
\label{eq:glcond_T}
\end{equation}
in which the second term of the right-hand side gives a numerically minor contribution to the sum rule. In the actual calculation, the Gell-Mann$-$Oakes$-$Renner (GOR) relation is used for evaluating the vacuum quark condensate,
$m_q\bra\bar{q}q\ket=-(0.11\,\mathrm{GeV})^4$, while the charmonium sum rules~\cite{Ioffe:2005ym} constrain the vacuum gluon condensate as $\glcond=0.005\pm0.004\,\mathrm{GeV}^4$, with large uncertainty. In practice, this uncertainty is not prohibitive since the overall correction induced by $c_2$ in Eq. (\ref{1momsr}) is small compared to the dominant term proportional to $s_0^2(T)$.

Apart from the in-medium modifications of the scalar condensates, new spin-dependent operators appear in the OPE because of broken Lorentz invariance in the heat bath. Such operators are classified by their canonical dimension and twist ($\tau=\text{dimension}-\text{spin}$). The $c_2(T)$ in the sum rule for the first moment now includes the contribution from the twist-2 operator $\mathcal{ST}\,\bar{q}\gamma_\mu D_\nu q$, where the symbol $\mathcal{ST}$ makes the operator symmetric and traceless with respect to its Lorentz indices. The pion matrix element of this operator in Eq. (\ref{eq:softpion}) is evaluated using its relation to the quark distribution function in the pion~\cite{Hatsuda:1992bv}. In the actual calculation, this contribution to $c_2(T)$ is about three times smaller than the leading $T$-dependent correction to the quark condensate in Eq. (\ref{eq:LOqucondT}).

\section{Phenomenology at low $T$}

The mixing of vector and axial-vector correlation functions at finite temperature is introduced following Ref. \cite{Dey:1990ba} and translates correspondingly to the spectral functions:  
\begin{eqnarray}
    R_V(s,T)&=&\left[1-\epsilon(T)\right]\,R_V(s,0) + \epsilon(T)\,R_A(s,0)~,\nonumber\\
     R_A(s,T)&=& \epsilon(T)\,R_V(s,0) + \left[1-\epsilon(T)\right]\,R_A(s,0).
        \label{eq:V-Amixing}
\end{eqnarray}
The mixing parameter $\epsilon$ is given by the thermal pion loop integral
\begin{equation}
    \epsilon(T)=\frac{2}{f_\pi^2}\int\frac{\ud^3p}{(2\pi)^3}\frac{1}{E(e^{E/T}-1)}~.
\end{equation}
In the chiral limit ($m_\pi\to0$) it reduces to $\epsilon=T^2/(6f_\pi^2)$. At the temperature $T_d$ where $\epsilon\simeq0.5$, the vector and axial-vector spectral functions are maximally mixed and become degenerate, $R_V(s,T)=R_A(s,T)$. The temperature at which this degeneracy takes place,
$T_d = \sqrt{3}\,f_\pi\simeq 151$ MeV (with $f_\pi = 87$ MeV in the chiral limit), is not far from the characteristic temperature commonly associated with the chiral crossover transition ($T_c = 170 - 190$ MeV) \cite{Cheng:2006qk, Aoki:2006br}. Of course, as one approaches this temperature range,
terms of higher order in $T^2$ become important and need to be taken into account.

It was pointed out in Ref.~\cite{Dey:1990ba} that the vector meson mass changes weakly at low
temperature (i.e., at order $T^2$). We demonstrate this feature next by performing the FESRs introduced in the previous sections.

\subsection{Spectral functions with zero width}
It is instructive to test the FESR for the vector and axial-vector spectral functions with zero width, 
the limiting situation realized in the large $N_c$ limit of QCD:
\begin{equation}
 \begin{split}
    R_V(s,T=0)&=12\pi^2\,f_V^2\,m_V^2\,\delta(s-m_V^2)~,\\
    R_A(s,T=0)&=12\pi^2\,f_A^2\,m_A^2\,\delta(s-m_A^2)~.
 \label{eq:spftn0}
 \end{split}
\end{equation}
The vector and axial-vector couplings, $f_{V/A}$, are determined by
\begin{equation}
    f_V^2\,m_V^2=2f_\pi^2,\qquad
    f_A^2\,m_A^2\simeq f_\pi^2+\frac{s_A-s_V}{8\pi^2}~.
\label{eq:fVA}
\end{equation}
The first part is just the KSRF relation; the second one is derived by applying the KSRF relation
to the first Weinberg sum rule in Eq.~(\ref{eq:largeNWSR}) and adding the correction coming from the difference between $s_A$ and $s_V$. Then the lowest two moments of the vector sum rule are: 
\begin{eqnarray}
    12\pi^2\,f_V^2\,m_V^2\left[1-\epsilon(T)\right]&=&\frac{3}{2}\,s_V(T)-12\pi^2f_\pi^2\,\epsilon(T)~,\nonumber\\
      \label{eq:chiLVsr0}\\
    12\pi^2\,f_V^2\,m_V^4\left[1-\epsilon(T)\right]&=&\frac{3}{4}\,s_V^2(T)-c_2(T)~,
      \label{eq:chiLVsr1}
\end{eqnarray}
where the $a_1$ meson contribution has been omitted because in the actual calculation, $\sqrt{s_V}$ is found to be located at  $s_V\leq m_A^2$ so that the $a_1$ pole does not contribute to the integrals of the spectral moments.

With this setup, the continuum threshold in the vector channel is evaluated to be $\sqrt{s_V}\simeq1.1~\mathrm{GeV}$ at $T=0$  from the sum rules, using $f_\pi\simeq87~\mathrm{MeV}$ in the chiral limit and $m_V=770~\mathrm{MeV}$. Note that $\sqrt{s_V}$ turns out to be perfectly close to the chiral scale, $4\pi\,f_\pi$. 

\begin{figure}[ht]
    \centering
    \includegraphics[width=8cm]{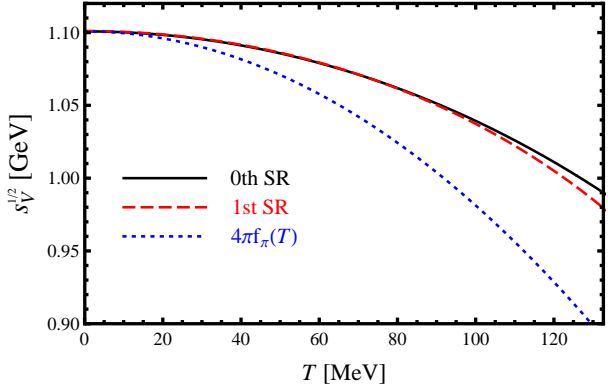}
    \caption{(Color online) Continuum threshold $\sqrt{s_V}$ in the vector meson channel as a function of $T$ obtained from the sum rules for the 0th (black solid line) and 1st (red dashed line) moment of the spectral function. The $T$ dependence of chiral scale $4\pi f_\pi(T)$ (blue dotted line) is also displayed.}
\label{fig:chirho}
\end{figure}

The finite temperature behavior of $\sqrt{s_V(T)}$ extracted from the sum rules to leading order in $\epsilon$ is shown in Fig.~\ref{fig:chirho}. The sum rules for the lowest two moments give consistent results at low temperature. However, although the identification $\sqrt{s_V}=4\pi f_\pi$ emerges naturally from Eq.~(\ref{eq:chiLVsr0}) at $T=0$, the temperature dependence of $\sqrt{s_V}$ evolves
more slowly than the canonical $f_\pi(T)=\left[1-\epsilon(T)/2\right]f_\pi(T=0)$ found in chiral perturbation
theory~\cite{Gasser:1986vb}. 

Defining an average mass $\bar{m}$ by the normalized first moment of the spectral function,
\begin{equation}
    \bar{m}^2\equiv\frac{\int\ud{s}\,sR}{\int\ud{s}\,R}~,
\label{eq:avmass}
\end{equation}
we confirm that the $\rho$ meson mass remains unchanged for $T\neq0$ and stays at the vacuum pole position, $\bar{m}_V=770~\mathrm{MeV}$, to leading order in $\epsilon$. This is consistent with the statement \cite{Dey:1990ba,Eletsky:1994rp} that the vector meson pole position remains unchanged at order $T^2$.

\begin{figure}[ht]
    \centering
    \includegraphics[width=8cm]{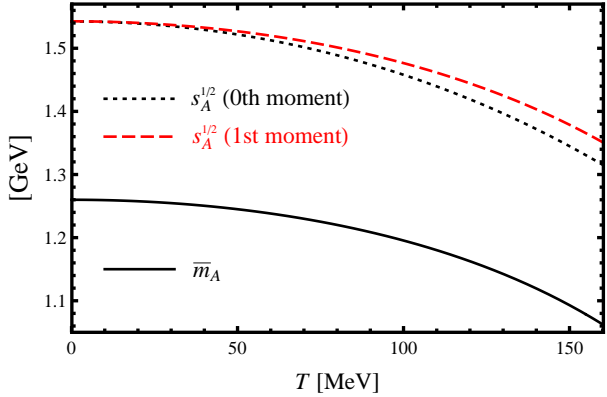}
    \caption{(Color online) Axial-vector continuum threshold $\sqrt{s_A}$ as a function of $T$ obtained from the sum rules for the 0th (black solid line) and 1st (red dashed line) moment of the $a_1$ spectral function. In contrast to the $\rho$ meson, the $a_1$-meson mass (black dotted line) decreases with rising temperature.}
\label{fig:chia1}
\end{figure}
In contrast to the $\rho$ meson case in which the $a_1$ pole does not contribute to the integrals of the spectral moments (since $s_V\leq m_A^2$), the left-hand side of the sum rules in the axial-vector channel receives contributions from both $\rho$ and $a_1$ poles. By using Eq.~(\ref{eq:fVA}) the sum rule for the
first moment of the $a_1$ spectral function becomes 
\begin{equation}
    12\pi^2\left[f_A^2\,m_A^4(1-\epsilon)+f_V^2\,m_V^4\,\epsilon\right]=\frac{3}{4}s_A^2-c_2~.
\label{eq:chiLAsr1}
\end{equation}
With $m_A=1.26~\mathrm{GeV}$, this gives $\sqrt{s_A}\simeq1.54~\mathrm{GeV}$ at $T=0$ and thus $f_A^2\simeq0.014$, which agrees with the empirical values of $f_A^2$ in Ref.~\cite{Sakurai:1980js}. As seen in Fig.~\ref{fig:chia1}, the average axial-vector mass $\bar{m}_A$ defined by Eq.~(\ref{eq:avmass}) decreases in parallel with the continuum threshold $\sqrt{s_A}$ as the temperature rises.
One observes that the $a_1$ mass $\bar{m}_A\simeq1.09~\mathrm{GeV}$ does not yet approach the $\rho$ mass $\bar{m}_V=770~\mathrm{MeV}$ at temperatures around  $T_d\simeq151~\mathrm{MeV}$, where $\epsilon=0.5$, suggesting that higher order effects in $\epsilon$ become important as one gets closer to the critical region for chiral restoration.

\subsection{Realistic spectral functions with finite width}

\begin{figure}[ht]
    \centering
    \includegraphics[width=8cm]{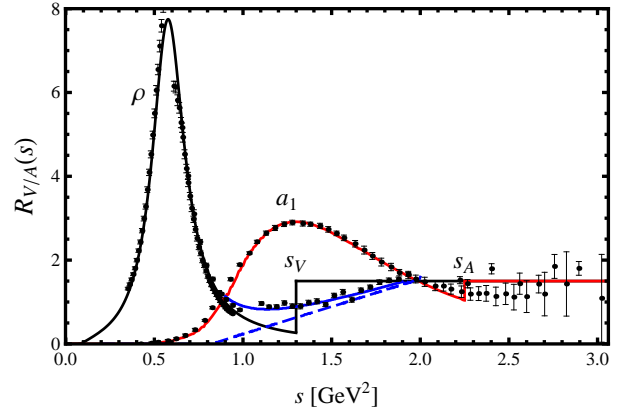}
    \caption{(Color online) Vector (black curve) and axial-vector (red curve) spectral distributions in vacuum, compared to $e^+e^-\to{n}\pi$ data with $n$ even~\cite{Aloisio:2004bu,Dolinsky:1991vq} and data from hadronic $\tau$ decays~\cite{Barate:1998uf,Ackerstaff:1998yj}. Here $s_V$ and $s_A$ stand for the continuum thresholds in vector and axial-vector channels, respectively. The ramping function (blue dashed line) shows an example of smooth threshold modeling (see text and appendix).}
\label{fig:specftn}
\end{figure}

In this section we study realistic empirical spectral functions with finite widths as displayed in Fig.~\ref{fig:specftn}. The continuum is still described by a step function. In previous
work~\cite{Kwon:2008vq}, we have shown that, both in the vacuum and at finite baryon density,  the detailed choice for modeling the continuum threshold is not decisive for the outcome of the sum-rule analysis. In practice, the continuum threshold can be given an improved description using a 
ramping function, resulting in the blue solid curve in Fig.~\ref{fig:specftn}, which reproduces the experimental data very well. We show that for the finite-temperature sum rules considered here, the results are also consistently stable, independent of the detailed threshold modeling, as long as the slope of the ramping into the continuum is chosen sufficiently large (see Appendix~\ref{appx2}).

\begin{figure}[ht]
    \centering
    \includegraphics[width=8cm]{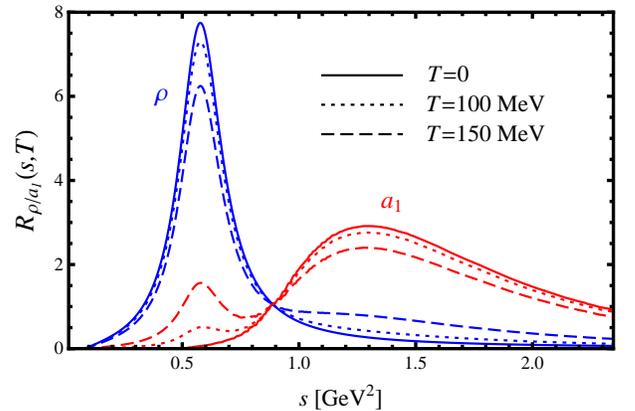}
    \caption{(Color online) The $\rho$ and $a_1$ meson spectra at finite temperature.}
\label{fig:specFT}
\end{figure}
Figure \ref{fig:specFT} exhibits the temperature-dependent $\rho$ and $a_1$ spectra generated by Eq.~(\ref{eq:V-Amixing}) with physical values of the pion mass and decay constant, $m_\pi=139.6~\mathrm{MeV}$ and $f_\pi=92.4~\mathrm{MeV}$. The pole position of the vector meson resonance stays at its vacuum mass. However, when examining the average mass defined by Eq.~(\ref{eq:avmass}),
\begin{equation}
    \bar{m}_V^2(T) = \frac{\frac{c_0}{2}s^2_V(T)-c_2(T)}{c_0\,s_V(T) +c_1-12 \pi^2 f_\pi^2\, \epsilon(T)}~,
\label{eq:avVmass}
\end{equation}
the broad widths of the $\rho$ and $a_1$ spectra affect the $\rho$ mass distribution.
\begin{table}[ht]
\centering
\begin{tabular}{|c|c|c|}
  \hline
  $T$ (MeV) & $\bar{m}_V$ (MeV) & $\sqrt{s_V}$ (GeV) \\
  \hline
    $0$ & ~$791\pm2$\,\, & ~$1.139\pm0.007$\,\, \\
   $50$ & ~$792\pm2$\,\, & ~$1.138\pm0.007$\,\, \\
   $100$ & ~$800\pm4$\,\, & ~$1.133\pm0.008$\,\, \\
  $120$ & ~$807\pm5$\,\, & ~$1.128\pm0.009$\,\, \\
  \hline
\end{tabular}
    \caption{Average vector mass and continuum threshold at various $T$ from Eqs.~(\ref{eq:avVmass}) and (\ref{1momsr}). The errors include uncertainties in the value of $\alpha_s$ (entering the NLO perturbative QCD corrections) and of the gluon condensate. }
\label{tab:avVmass}
\end{table}
From Table~{\ref{tab:avVmass}}, it is apparent that the average vector mass tends to slightly move upward with temperature, picking up weight from the mixing with the $a_1$, while $\sqrt{s_V}$ moves downward. When using the tightly constrained vacuum spectral functions as input at $T=0$, it turns out that there can be a small (<5\%) mismatch between the left-hand and the right-hand sides of the sum rules.  The error assignments in the values of $\sqrt{s_V}$ from the sum rule for the first spectral moment take this small uncertainty into account in Table~{\ref{tab:avVmass}}. In practice, the errors in
Table~\ref{tab:avVmass} come primarily from the uncertainties of the gluon vacuum condensate and the running strong coupling $\alpha_s(s)$. 

As the temperature increases the continuum onset scale $s_V(T)$ shows evidentally a tendency to decrease. At temperatures $T > 140$ MeV, however, solutions for $\sqrt{s_V}$ cease to exist,
and hence the average mass at that high $T$ cannot be determined via Eq.~(\ref{eq:avVmass}). This is partly a consequence of the restrictions imposed by the treatment to leading order in $\epsilon(T)$. Even more so, it is related to the fact that, with realistic spectral functions, the broad $a_1$ distribution does not permit a separation of scales between low- and high-energy parts of the spectrum. The increased mixing of the $a_1$  into the $\rho$ spectrum at high temperature enhances this effect.
Thus, with realistic spectral distributions, the FESR approach has its applicability limited to
the range well below the ``melting'' temperature of the resonances.

As mentioned in the introduction, the difference between vector and axial-vector correlators serves as an order parameter of spontaneous chiral symmetry breaking and restoration. Direct substraction of the FESRs in the axial-vector and vector channels gives
\begin{equation}
\begin{split}
    \int^{s_V(T)}_0\ud s\,R_V&(s,T)-\int^{s_A(T)}_0\ud s\,R_A(s,T)\\
        =c_0&\left[s_V(T)-s_A(T)\right]+12\pi^2f_\pi^2\left[1-2\epsilon(T)\right]~,\\
    \int^{s_V(T)}_0\ud s\,s\,R_V&(s,T)-\int^{s_A(T)}_0\ud s\,s\,R_A(s,T)\\
        &=\frac{c_0}{2}\left[s_V^2(T)-s_A^2(T)\right]~.
\label{eq:V-Asr}
\end{split}
\end{equation}
Keeping the leading order mixing ansatz (\ref{eq:V-Amixing}) for the spectral functions, one finds $\sqrt{s_V}\simeq1.12$ GeV and
$\sqrt{s_A}\simeq1.49$ GeV at $T=0$ from Eqs. (\ref{eq:V-Asr}). Note that $\sqrt{s_V}$ is again close to the chiral scale $4\pi f_\pi$. It is important to realize that $s_A$, as it results from the consistency conditions of the sum-rule analysis, is different from (i.e., significantly larger than) $s_V$.  In fact, the $a_1$ mass itself is comparable to $4\pi f_\pi$ so that the continuum threshold in the axial-vector channel must necessarily be located at a higher scale than $s_V$ once the large $a_1$ width is taken into account.  Assuming $s_A=s_V \equiv s_0$ as in the schematic Weinberg sum rules (\ref{eq:largeNWSR}), one finds that the leading order temperature dependence, $1-2\epsilon$, just drops out in Eq.~(\ref{eq:V-Asr}) and yields a trivial result,
\begin{equation}
\begin{split}
    \int^{s_0}_0\ud s\,\left[R_V(s,0)-R_A(s,0)\right]&=12\pi^2f_\pi^2~,\\
    \int^{s_0}_0\ud s\,s\left[R_V(s,0)-R_A(s,0)\right]&=0~,
\end{split}
\end{equation}
independent of temperature.

\section{Higher order corrections}
\subsection{$T^4$ corrections}
Effects at order $T^4$ have been treated systematically in Ref.~\cite{Eletsky:1994rp}. It is shown there that these corrections amount to replacing the mixing parameter $\epsilon$ by
$\epsilon\,\to\,\epsilon(1-\epsilon/2)$. The temperature $T_d$ at which the $\rho$ and $a_1$ spectral distributions become degenerate is now shifted upward to $T_d = \sqrt{6} f_\pi$.

The $T^4$ correction incorporates interacting pions  with nonvanishing momentum in the heat bath, which contribute to the Lorentz noninvariant part of the current correlation
function. According to the Ref.~\cite{Eletsky:1994rp}, this contribution can be expressed in the frame with
$\mathbf{q}=0$ as follows:
\begin{equation}
\begin{split}
    \Pi_V(q^2 = -Q^2,T)~&=~ \Pi_V(-Q^2,T=0)\\
            \quad-~\epsilon\left(1-\frac{\epsilon}{2}\right)&\left[\Pi_V(-Q^2,T=0)-\Pi_A(-Q^2,T=0)\right]\\
            +&\frac{4\pi^2\,M_2}{15 \,Q^2}\,T^4~,
\label{eq:corT^4}
\end{split}
\end{equation}
where $M_2$ is the first moment of quark distributions in the pion:
\begin{equation}
    M_2=\frac{1}{2}\int^1_0\ud x\,x\left[v(x)+2s(x)\right]~,
\end{equation}
with valence and sea quark distributions, $v(x)$ and $s(x)$, respectively. We use the value $M_2 \simeq 0.12$ as discussed in \cite{Eletsky:1994rp}.

By transferring Eq.~(\ref{eq:corT^4}) to the FESR, it turns out that Eq.~(\ref{1momsr})
is modified by the new term of order $T^4$ as follows:
\begin{equation}
    \int^{s_V(T)}_0\ud s\,s\,R_V(s,T) -\frac{16\pi^4M_2}{5}\,T^4=\frac{c_0}{2} s^2_V(T)-c_2(T)~,
        \label{1momsrc}
\end{equation}
while the denominator of Eq.~(\ref{eq:avVmass}) receives a small correction from the replacement
$\epsilon\,\to\,\epsilon(1-\epsilon/2)$. 

On the OPE side, the $T$ dependence of the quark condensates is also improved up to order $T^4$. 
Chiral perturbation theory gives~\cite{Gerber:1988tt}
\begin{equation}
    \qucond_T=\qucond_0\left(1-\frac{3}{4}\epsilon-\frac{3}{32}\epsilon^2\right).
\end{equation}
However, this correction is numerically small in the actual calculation.

The substitution of Eq.~(\ref{1momsrc}) in the numerator of Eq.~(\ref{eq:avVmass}) introduces a relatively small negative mass shift,
\begin{equation}
   \delta \bar{m}_V^2 = -\frac{16\pi^4M_2\,T^4}{5 \int^{s_V(T)}_0\ud s\,R_V(s,T)}~, 
\end{equation}
consistent with the findings of Ref.~\cite{Eletsky:1994rp}, 
which compensates for the increase of the vector average mass by the finite width effect. For instance the average mass at $T=100$ MeV is now obtained as $\bar{m}_V\simeq 798\pm4$ MeV, slightly less  than in Table~\ref{tab:avVmass} but only marginally different. 
\subsection{Massive states}
For higher temperatures the contributions from massive excitations such as $K$ and $\eta$ in Eq.~(\ref{eq:TherAV}) need to be considered. In the $T$ dependence of the quark and gluon condensates, these contributions are included as in Ref.~\cite{Hatsuda:1992bv}:
\begin{equation}
\begin{split}
    \frac{\qucond_T}{\qucond_0}&=1-\frac{T^2}{8f_\pi^2}\left[B_1\left(\frac{m_\pi}{T}\right)+\frac{7}{9}B_1\left(\frac{m_K}{T}\right)\right],\\
    \glcond_T&=\glcond_0-\frac{T^2}{9}\left[m_\pi^2B_1\left(\frac{m_\pi}{T}\right)\right.\\
             &\qquad\qquad\qquad\quad+\left.\frac{5}{3}m_K^2B_1\left(\frac{m_K}{T}\right)\right].
\end{split}
\end{equation}
In practice, the $T$ dependence of the gluon condensate is just a few percent of its vacuum value and negligible. These numerically minor corrections in the OPE are of only little significance to the sum-rule analysis.

\section{Spectral functions from an effective field theory}

So far we have used empirical input for the spectral distributions at zero temperature. 
In this section, we illustrate how the sum rule works for the example of a
chiral effective field theory in which the vector meson spectral distribution can be
explicitly calculated and tested with respect to its sum-rule consistency. 
The example chosen is the chiral Lagrangian based on generalized hidden local 
symmetry (GHLS)~\cite{ghls}, which explicitly includes the axial-vector meson in addition to 
pion and vector meson. Details of the formalism at one loop are found in Ref.~\cite{our}.

The parity mixing in the $\rho$ meson spectrum at finite temprature is generated from 
a process in which $\pi$ and $a_1$ mesons circulate in a loop attached to the $\rho$ meson,
with the pion coming from the heat bath. In this approach one does not have to rely on 
equations such as Eqs.~(\ref{eq:V-Amixing}) or (\ref{eq:corT^4}).

\begin{figure}[ht]
    \centering
    \includegraphics[width=8cm]{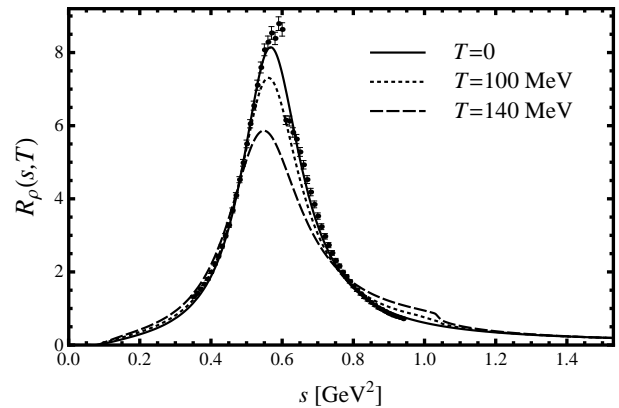}
    \caption{The $\rho$ meson resonance at finite temperature calculated in the GHLS model.}
\label{fig:ghls}
\end{figure}

Figure~\ref{fig:ghls} shows the vector spectral function of the GHLS approach at
several temperatures.
The spectrum exhibits the usual resonant peak, the width of which is governed by the 
$\rho \rightarrow 2\pi$ decay. The $\rho$-$a_1$ mixing involves mechanisms in which the
energy of the timelike 
$\rho$ meson splits into two branches corresponding to the processes 
$\rho + \pi \to a_1$ and $\rho \to a_1 + \pi$, with thresholds 
$\sqrt{s} = m_{a_1} - m_\pi$ and $\sqrt{s} = m_{a_1} + m_\pi$.
This produces the threshold effects seen as a shoulder at 
$\sqrt{s}=m_{a_1}- m_\pi$ (and a bump above $\sqrt{s}=m_{a_1} + m_\pi$, not
seen in the figure). The height of the $\rho$ spectrum gets reduced with increasing temperature,
whereas the $a_1$-meson contribution is enhanced via the mixing effect.

The average vector mass defined in Eq.~(\ref{eq:avmass}) and the continuum
threshold in the vector channel as a function of  temperature are summarized
in Table~\ref{tab:sthHLS}.
\begin{table}[ht]
\centering
\begin{tabular}{|c|c|c|}
  \hline
  $T$ (MeV) & $\bar{m}_V$ (MeV) & $\sqrt{s_V}$ (GeV) \\
  \hline
    $0$ & ~$787\pm2$\,\, & ~$1.126\pm0.007$\,\, \\
   $40$ & ~$787\pm2$\,\, & ~$1.126\pm0.007$\,\, \\
   $80$ & ~$787\pm2$\,\, & ~$1.122\pm0.007$\,\, \\
  $120$ & ~$786\pm3$\,\, & ~$1.111\pm0.008$\,\, \\
  $140$ & ~$786\pm3$\,\, & ~$1.102\pm0.008$\,\, \\
  \hline
\end{tabular}
\caption{The average $\rho$-meson mass and the continuum
threshold $\sqrt{s_V}$ at various temperatures, following from the FESR consistency test of the 
spectral function calculated in the GHLS approach \cite{our}.}
\label{tab:sthHLS}
\end{table}
One observes again that $\bar{m}_V$ stays unchanged from its vacuum value over the whole
temperature range $T < 140$ MeV. The continuum threshold scale $\sqrt{s_V}$ shows a systematic decrease toward higher temperature, but at a rate considerably smaller than the expected
behavior of the chiral order parameter, $f_\pi(T)$.

\section{Conclusions}

In this work we have constructed FESRs in order to study the behavior of $\rho$ and $a_1$ mesons as well as their mixing at finite temperature, with the aim of exploring the pattern of chiral symmetry restoration.  The sum rules for the lowest two spectral moments of vector and axial-vector spectral functions involve only the leading QCD condensates as corrections. With inclusion of perturbative QCD terms up to order $\alpha_s^3$, these sum rules permit a reliable  quantitative analysis, unaffected by the large uncertainties from condensates of higher dimension such as the four-quark condensates.

The leading temperature corrections involve thermal pion loops. To order $T^2$, the temperature dependence is generated entirely by the $\rho - a_1$ mixing effect caused by the nonvanishing $\rho\,\pi \,a_1$ coupling in the presence of a pionic heat bath, as pointed out previously by Dey, Eletsky, and Ioffe  \cite{Dey:1990ba}. To this order, there is no shift of the pole mass in the thermal vector meson propagator.  Next-to-leading effects of order $T^4$ involve $\pi\pi$ intermediate states and can lead to  moderate mass shifts. 

In the large-$N_c$ limit the $\rho$ and $a_1$ widths vanish. In this limit, schematic distributions with $\delta$ function resonances and a step-function parametrization of the high-energy continuum, when inserted in the FESR analysis, reproduce the well-known current algebra and chiral sum rules. 
As an interesting feature, one finds that at zero temperature the continuum threshold $s_V$ in the vector channel is identified with the scale characteristic of spontaneous chiral symmetry breaking:
$\sqrt{s_V} = 4\pi\,f_\pi$. Above this scale, chiral symmetry is restored in its Wigner-Weyl realization. In the resonance region below this scale, the symmetry is in the spontaneously broken Nambu-Goldstone realization.

When realistic spectral functions with large widths are implemented, this ``clean'' separation of scales 
is no longer rigorously maintained, but the FESR analysis is still useful, with the pole mass in the vector correlator now replaced by the normalized first moment of the spectral distribution. Within its range of applicability (up to temperatures of about 140 MeV), the sum-rule analysis consistently shows an  almost constant behavior of this average vector meson  spectral mass. 
The primary temperature dependence of the spectral function comes from $\rho$-$a_1$ mixing in the thermal pionic medium. The $a_1$ mass, again identified with the normalized first moment of the
axial-vector spectral distribution, decreases with rising temperature. This indicates the expected tendency of the $\rho$ and $a_1$ spectra becoming identical (degenerate) when chiral symmetry is restored. However, the present analysis does not support the BR scaling hypothesis of a dropping $\rho$ meson mass, at least not up to temperatures $T\sim140$ MeV, which are not far from the chiral crossover transition temperature, $170-190$ MeV.

The continuum threshold scale $s_V$ in the vector channel (even when smoothed by a ramping function) systematically moves downward in energy as the temperature increases. This feature is
observed in all cases studied. However, while the identification of $\sqrt{s_V}$ with the chiral scale
$4\pi\,f_\pi$ emerges naturally at $T = 0$, the downward evolution with temperature of $\sqrt{s_V}$ is  
significantly slower than that of $f_\pi(T)$ deduced from chiral perturbation theory.

\section*{ACKNOWLEDGEMENTS}
This work has been supported in part by BMBF, GSI, and the DFG Cluster of Excellence Origin and Structure of the Universe. Two of us (WW and CS) thank the organizers of the program `New Frontiers in QCD' for their kind hospitality at the Yukawa Institute of Theoretical Physics, Kyoto, where this article was finalized. YK is grateful to S. H. Lee and S. i. Nam for useful discussions.
CS acknowledges parital support by the Hessian
LOEWE initiative through the Helmholtz International
Center for FAIR (HIC for FAIR).

\begin{appendix}
\section{Continuum threshold modeling}
\label{appx2}

Here we test the reliability of the continuum threshold parametrization by a schematic step function. 
Such a test can be performed by replacing the step function with a ramp function to yield a smooth transition between resonance and continuum regions, as follows:
\begin{equation}
     R(s)=R_{res}(s)\,\Theta(s_2-s)+R_c(s)\,W(s)~,
\end{equation}
where the weight function, $W(s)$, is defined as
\begin{equation}
  W(x)=\left\{\begin{array}{cl}
           \vspace{2mm}
          0 & \text{ for } x\leq s_1\\
           \vspace{1.5mm}
          \displaystyle\frac{x-s_1}{s_2-s_1} & \text{ for }s_1\leq x\leq s_2\\
          1 & \text{ for } x\geq s_2~.
        \end{array}\right.
  \label{eq:Wofs}
\end{equation}
The step function behavior is recovered for $W(x)$ in the limit $s_1\rightarrow s_2$.
\begin{figure}[ht]
   \includegraphics[width=8.5cm]{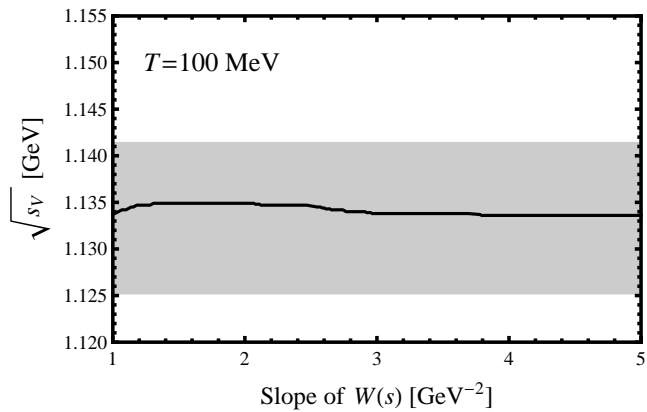}
   \caption{Dependence of $\sqrt{s_V}$ [determined from Eqs.~(\ref{app:0thsr})-(\ref{app:s_v})] on the slope
    $(s_2-s_1)^{-1}$ of  the ramp function $W(s)$ describing the onset of the continuum for the example of $T=100~\mathrm{MeV}$.
    The grey band indicates the uncertainty range of the result obtained with step-function parametrization of the continuum.}
   \label{app:ramp}
\end{figure}

By using the function $W(s)$, the modified sum rules for the lowest two moments of the spectrum $R(s)$ become
\begin{equation}
\begin{split}
   \int^{s_2}_0\ud s\,R_{res}(s)&=s_2\left(c_0+\frac{3}{2}\varepsilon_0\right)+c_1-12\pi^2\Pi(0)\\
                            &-\left[c_0-R_{res}(s_2)\right]\int^{s_2}_{s_1}\ud s\,W(s)~,
              \label{app:0thsr} 
\end{split}
\end{equation}
              
\begin{equation}
\begin{split}
   \int^{s_2}_0\ud s\,sR_{res}(s)&=\frac{s^2_2}{2}\left(c_0+\frac{3}{2}\varepsilon_1\right)-c_2\\
                             &-\left[c_0-R_{res}(s_2)\right]\int^{s_2}_{s_1}\ud s\,sW(s)~.
              \label{app:1stsr}
\end{split}
\end{equation}
Sets of intervals $[s_1, s_2]$ are then determined so as to satisfy both sum rules (\ref{app:0thsr}, \ref{app:1stsr}),
and the scale $s_V$, defined by
\begin{equation}
   s_V=\frac{s_1+s_2}{2}~,
   \label{app:s_v}
\end{equation}
is now introduced to characterize the continuum threshold.
In any temperature region where the sum rules are valid the present analysis does not depend sensitively on details of the threshold modeling. For an example of the $\rho$ meson channel, it is demonstrated in Fig.~\ref{app:ramp} that the resulting $\sqrt{s_V}$ at finite temperature is
stable with respect to variations in the slope $(s_2-s_1)^{-1}$ of the ramp function $W(s)$. In this test the
uncertainties of $\alpha_s(Q^2)$ and of the gluon condensate have been disregarded  for simplicity.
\vspace{0.5cm}

\end{appendix}



\end{document}